\documentstyle[12pt]{article}

\newcommand{\newsection}{    
\setcounter{equation}{0}
\section}
\newcommand{\tr}[1]{\,{\rm tr}\,#1\,}
\def\e{{\,\rm e}\,}
\def\pint{\int\hspace{-1.07em}\not\hspace{0.6em}}

\def\eop{\vspace*{\fill}\pagebreak}
\def\be{\begin{equation}}
\def\ee{\end{equation}}
\def\bea{\begin{eqnarray}}
\def\eea{\end{eqnarray}}

\newcommand{\rf}[1]{(\ref{#1})}
\newcommand{\eq}[1]{Eq.~(\ref{#1})}

\def\bb{\bar{\beta}}
\def\bm{\bar{m}}

\def\l{\lambda}

\newcommand{\ie}{{\it i.e.}\ }

\newcommand{\etal}{{\it et al\/.\ }}

\newcommand{\half}{{\textstyle{1\over 2}}}
\renewcommand{\d}{{{\partial}}}

\newcommand{\ra}{\rightarrow}
\newcommand{\fr}[2]{{\textstyle {#1 \over #2}}}
\begin{document}

\begin{flushright}
ITEP-YM-6-92 \\ August, 1992
\end{flushright}

\begin{center}
{\LARGE Large-N Reduction, Master Field and \\
\vspace{0.6cm} Loop Equations in Kazakov--Migdal Model}
\end{center}
\vspace{1.5cm}
\begin{center} {\large Yu.\
Makeenko}\footnote{E--mail: \ makeenko@itep.msk.su \ \ / \ \
makeenko@desyvax.bitnet \ \ / \ \ makeenko@nbivax.nbi.dk } \\ \mbox{} \\
{\it Institute of Theoretical and Experimental Physics} \\ {\it
B.Cheremuskinskaya 25, 117259 Moscow, Russia}
\end{center}

\vspace{1cm}

\begin{abstract}
I study the large-$N$ reduction {\it a la} Eguchi--Kawai in the
Kazakov--Migdal lattice gauge model. I show that both quenching and
twisting prescriptions lead to the coordinate-independent master field.  I
discuss properties of loop averages in reduced as well as unreduced models
and demonstrate those coincide in the large mass expansion. I derive loop
equations for the Kazakov--Migdal model at large $N$ and show they are
reduced for the quadratic potential to a closed set of two equations.
I find an exact strong coupling solution of these equations for any $D$
and extend the result to a more general interacting potential.
  \end{abstract}


\eop

\newsection {Introduction}

The purpose of this paper is a further study of the
Kazakov--Migdal lattice gauge theory~\cite{KM92} which is defined by
the partition function
 \be
Z_{KM}=\int \prod_{x,\mu} dU_{\mu}(x) \prod_x d\Phi(x)
\e^{\sum_x N \tr{\left(-V[\Phi(x)]+
\sum_{\mu}\Phi(x)U_\mu(x)\Phi(x+\mu)U_\mu^\dagger(x)\right)}}\;.
\label{partition}
\ee
Here the field $\Phi(x)$ takes values in the adjoint representation of the
gauge group $SU(N)$ and the link variable $U_\mu(x)$ is an element of the
group.  The lowest coefficient of the expansion of the potential $V[\Phi]$
in the lattice spacing, $a$,
\be
V[\Phi]=m_0^2\Phi^2+ \ldots
\label{potential}
\ee
is identified with a bare mass parameter while the others play the role of
couplings of self-interaction of the field $\Phi$.

The recent interest in the model \rf{partition} is due to the following
reasons. As it is proposed in Ref.~\cite{KM92} the model \rf{partition}
induces QCD in the continuum limit which should be obtained, as usually
in lattice gauge theories, in the vicinity of a second order phase
transition. This limit should be reached by approaching the couplings of the
potential~\rf{potential} to the corresponding critical values. As was
proposed in Ref.~\cite{KM92} this critical point should be identified with
the one separating the {\it strong\/} and {\it weak\/} coupling phases of
the model~\rf{partition}. On the other hand the model~\rf{partition} is
potentially solvable in the large-$N$ limit with different distributions of
eigenvalues of the matrix $\Phi$ in the two phases~\cite{KM92,Mig92a}. To
obtain a confining continuum limit, one should approach the critical point
from the site of strong coupling phase while approaching from the weak
coupling phase seems to result in deconfining Higgs phase.

However, soon after this scenario of inducing QCD had been proposed, it was
pointed out by Kogan, Semenoff and Weiss~\cite{KSW92} that
the model~\rf{partition} possesses an extra local $Z_N$ symmetry which leads
in the strong coupling phase to {\it local\/} confinement when quarks can not
propagate even inside hadrons. To escape this physically unacceptable
picture, it was assumed this $Z_N$ symmetry to be spontaneously broken at
the point of the strong to weak coupling phase transition. A similar
scenario has been proposed by Khokhlachev and the author when normal
confinement restores after the large-$N$ phase transition with occurs
{\it before} the one associated with continuum limit (\ie within the strong
coupling phase in our terminology). This conjectured phase looks similar to
the weak coupling phase of standard lattice gauge theories.
To avoid terminological confusions I shall refer it as the {\it
intermediate} coupling phase.

The problem of whether the large-$N$ phase transition occurs according to
this scenario is a dynamical one and can be studied using standard
methods of lattice gauge theories. In particularly, the mean field method has
been applied in Ref.~\cite{KhM92}. The result for the case of the quadratic
potential $V[\Phi]$ in negative --- no first order phase transition occurs
in the strong coupling phase. This conclusion coincides with the one made on
the basis of the exact solution found for the quadratic potential by
Gross~\cite{Gro92}.

While the above scenario failed for the
quadratic potential for this reason, a possibility of an alternative
`stringy' continuum limit has been conjectured for this case by Kogan,
Morozov, Semenoff and Weiss~\cite{KMSW92} when the lattice spacing is taken
to be $N$-dependent and should approach zero simultaneously with
$N\ra\infty$ in a special way.  Such a `stringy' large-$N$ limit differs
from the 't Hooft topological expansion of QCD which is dictated by the
known dependence of the coupling constant on $N$ prescribed by asymptotic
freedom at small distances.  For this reason, all the standard large-$N$
technology like factorization, saddle point equations, {\it etc.} are not
applicable to the `stringy' large-$N$ limit. In particularly, the exact
$N=\infty$ solutions of Refs.~\cite{Mig92a,Gro92} can not be applied as
well.

The latter exact solutions have been obtained by solving the `master field
equation' derived by Migdal~\cite{Mig92a} under the assumption~\cite{KM92}
that the path integral over $\Phi(x)$ is saturated as $N\ra\infty$ by a
single $x$-independent saddle point configuration $\Phi_s$ --- the master
field. While it was pointed out~\cite{KM92} that such a master field does
not contradict to all our knowledge about the large-$N$ limit, a mechanism of
its appearance was mysterious.

In the present paper I consider this problem from the viewpoint of
the large-$N$ reduction which had been first advocated by Eguchi and
Kawai~\cite{EK82} for lattice gauge theories at $N=\infty$. The large-$N$
reduction states that the model on an infinite lattice is equivalent that
at one point (a plaquette in the case of lattice gauge theory) so that
the space-time degrees of freedom are eaten by the internal symmetry group.
My idea would be to identify the master field $\Phi_s$ with a the saddle
point configuration of a one-matrix model which appears after the reduction.
I consider both quenching~\cite{BHN82} and twisting~\cite{GAO83}
prescriptions of the large-$N$ reduction and argue in Section~2 that while
they correctly reproduce the perturbative expansion of the Kazakov--Migdal
model, any dependence on the quenched momenta can be absorbed by a
(nonperturbative) gauge transformation.

To justify this reduction procedure, I derive in Section~3 loop equations
for the Kazakov--Migdal model. I show that in addition to the {\it adjoint}
Wilson loop
\be
W_A(C)=\left\langle\frac{1}{N^2}
\left( \left| \tr{U(C)}\right|^2-1\right)\right\rangle
\label{adjloop}
\ee
where the average is understood with the same measure as in \eq{partition},
the objects of a new kind
\be
G(C_{xy})= \Big\langle \fr 1N \tr{\left(
\Phi(x)U (C_{xy}) \Phi(y) U(C_{yx}) \right)} \Big\rangle
\label{G}
\ee
emerges in the loop equations. This has a meaning of the average for an
open loop which is made gauge invariant by attaching scalar fields at the
ends. For the quadratic potential I obtain in the large-$N$ limit the closed
set of two equations.

In Section~4, I discuss properties of the loop averages both in reduced and
in unreduced cases and show explicitly of how they coincide to the leading
order of the large mass expansion. I speculate as well on the properties of
the intermediate phase and argue that it should resemble the standard Wilson
lattice gauge theory.

An {\it exact\/} solution of the loop equations in the strong coupling phase is
found in Section~4 for the quadratic potential at any number of dimensions
$D$.  An explicit formula for the average~\rf{G} is given by \eq{final} below.
While this solution agrees with that by Gross~\cite{Gro92}, I do not make
any assumptions about the master field to find it. Moreover it is a first
example of exact calculations of {\it extended\/} objects in Kazakov--Migdal
model.  I extend the solution to a more general potential \be
N\tr{V[\Phi]}= m_0^2 N \tr{\Phi^2}+N^2 f(\fr 1N \tr{\Phi^2}),
\label{general} \ee where $f$ is an arbitrary function, of the type studied
 recently in the matrix models~\cite{AGB92} and show that no large-$N$ phase
transition occurs in the strong coupling phase for this potential as well.

\newsection{Large-$N$ reduction}

\subsection{Scalar field}

The idea of large-$N$ reduction was putted forward by Eguchi and
Kawai~\cite{EK82} who showed the Wilson lattice gauge theory on a
$D$-dimensional hypercubic lattice to be equivalent at $N=\infty$ to the one
on a hypercube with periodic boundary conditions. This construction is based
on an extra $(Z_N)^D$-symmetry which the latter theory possesses to each
order of the strong coupling expansion but is broken in the weak coupling
region. To cure the construction at weak coupling, the quenching
prescription was proposed by Bhanot, Heller and Neuberger~\cite{BHN82} and
elaborated by many authors (for a review, see~\cite{Das87}). An elegant
alternative reduction procedure based on twisting prescription was advocated
by Gonzalez-Arroyo and Okawa~\cite{GAO83}. An extension of the quenched
Eguchi--Kawai model to the case of hermitian matrices was proposed by
Parisi~\cite{Par82} end elaborated by Gross and Kitazawa~\cite{GK82} while
that of the twisting prescription was advocated by Eguchi and
Nakayama~\cite{EN83} and has been discussed recently by Alvar\'{e}z-Gaume
and Barb\'{o}n~\cite{AGB92} in the context of $D>1$ strings.  Let me first
briefly review these results which allow to reduce the partition function of
self-interacting matrix scalar field on the infinite $D$-dimensional lattice
at $N=\infty$ to a hermitian one-matrix model in an external field.

For a pure scalar theory whose partition function is defined by the path
integral similar to \rf{partition} but without gauging:
 \be
Z=\int  \prod_x d\Phi(x)
\e^{\sum_x N \tr{\left(-V[\Phi(x)]+
\sum_{\mu}\Phi(x)\Phi(x+\mu)\right)}}\;,
\label{scalar}
\ee
the quenched momentum prescription is formulated as follows.  One
substitutes \be \Phi(x)\ra S(x) \Phi S^\dagger(x)
\label{substitution}
\ee
where
\be
[S(x)]_{ij} = \e^{ ik_i^\mu x_\mu} \delta_{ij}
\label{S}
\ee
is a unitary matrix which eats the coordinate dependence.
The averaging of a functional $F[\Phi(x)]$ with the same weight as in
\eq{scalar} can be calculated at $N=\infty$ by
\be
\Big\langle F[\Phi(x)]  \Big\rangle  \ra
\int_{-\pi}^\pi \prod_{\mu=1}^D \prod_{i=1}^N  \frac{dk_i^\mu}{2\pi}
\left\langle F[S(x)\Phi S^\dagger(x)]  \right\rangle_{Reduced}
\label{averages}
\ee
where the average on the r.h.s.\ is calculated for the {\it quenched reduced
model\/} defined by the partition function~\cite{Par82}
\be
Z_{QRM}=\int d\Phi \e^{-N\tr{V[\Phi]}+
N\sum_{ij} \left|\Phi_{ij}\right|^2
\left( D-\sum_\mu\cos{(k_i^\mu-k_j^\mu)}\right) }
\label{QRM}
\ee
which can be obtained from the one \rf{scalar} by the substitution
\rf{substitution}.

Since $N\ra\infty$ it is not necessary to integrate over the quenched
\sloppy momenta in \eq{averages}. The integral should be recovered if
$k_i^\mu$'s would be uniformly distributed in a $D$-dimensional hypercube.
Moreover, a similar property holds for the matrix integral over $\Phi$ as
well which can be substituted by its value at the saddle point configuration
$\Phi_s$:  \be \Big\langle F[\Phi(x)]  \Big\rangle  \ra F[S(x)\Phi_s
S^\dagger(x)] \;.  \label{masterfield} \ee This saddle point configuration
was referred as the {\it master field}~\cite{Wit80}.

An alternative reduction procedure is based on the twisting prescription.
One performs again the unitary transformation \rf{substitution} with the
matrices $S(x)$  being expressed via a set of $D$ (unitary)
$N\times N$ matrices $\Gamma_\mu$ by the path-dependent factors
\be
S(x)= P\prod_{l\in C_{x\infty}} \Gamma_\mu\; .
\ee
The path-ordered product in this formula runs over all links $l=(z,\mu)$
forming a path $C_{x\infty}$ from infinity to the point $x$. The matrices
$\Gamma_\mu$ are explicitly constructed in Ref.~\cite{GAO83} and commute by
\be
\Gamma_\mu \Gamma_\nu = Z_{\mu\nu} \Gamma_\nu \Gamma_\mu
\label{commutator}
\ee
with $Z_{\mu\nu}=Z_{\nu\mu}^\dagger$ being elements of $Z_N$.

Due to \eq{commutator}, changing the form of the path multiplies $S(x)$ by
the abelian factor
\be
Z(C)=\prod_{\Box\in S:\partial S=C} Z_{\mu\nu}(\Box)
\label{factor}
\ee
where $(\mu,\nu)$ is the orientation of the plaquette $\Box$.
The product runs over any surface spanned by the closed loop $C$ which is
obtained by passing the original path forward and the new path backward.
Due to the Bianchi identity
\be
\prod_{\Box\in cube} Z_{\mu\nu}(\Box)=1
\ee
where the product goes over six plaquettes forming a 3-dimensional cube on
the lattice, the product on the r.h.s.\ of \eq{factor} does not depend on
the form of the surface $S$ and is a functional of the loop $C$.

It is easy to see now that under this change of the path one gets
\be
[S(x)]_{ij}[S^\dagger(x)]_{kl}\ra
\left|Z(C)\right|^2[S(x)]_{ij}[S^\dagger(x)]_{kl}
\ee
and the path-dependence is cancelled because $\left|Z(C)\right|^2=1$.
This is a general property which holds for the twisting reduction
prescription of any even (\ie invariant under the center $Z_N$)
representation of $SU(N)$.

For definitiveness one can choose
\be
S(x)=\Gamma_1^{x_1}\Gamma_2^{x_2}\Gamma_3^{x_3}\Gamma_4^{x_4}
\ee
where the coordinates of the (lattice) vector $x_\mu$ are measured in the
lattice units.

For the twisting reduction prescription, \eq{averages} is valid providing
the average on the r.h.s.\ is calculated for the {\it twisted reduced
model\/} which is defined by the partition function~\cite{EN83}
\be
Z_{TRM}=\int d\Phi \e^{-N\tr{V[\Phi]}+
N\sum_\mu \tr{\Gamma_\mu \Phi \Gamma_\mu^\dagger \Phi}}\;.
\label{TRM}
\ee
Now the perturbation theory for the unreduced model \rf{scalar} is
recovered due to an explicit dependence of $\Gamma_\mu$ on momenta.

\subsection{Kazakov--Migdal model}

Analogous quenching and twisting reduction prescriptions hold for the
Kazakov--Migdal model as well.
To make consideration similar to that of the previous section, let us
introduce for the Itzykson--Zuber--Metha integral the notation
\be
I[\Phi,\Psi]\equiv\int dU \e^{N\tr{\Phi U \Psi U^\dagger}}
=\frac{\hbox{det}_{ij}\,\e^{\Phi_i\Psi_j}}{\Delta[\Phi]\Delta[\Psi]}
\label{IZ}
\ee
where $dU$ is the Haar measure on $SU(N)$ while
$\Phi_i$ and $\Psi_j$ stand for eigenvalues of the matrices
$\Phi$ and $\Psi$, respectively,
with $\Delta[\Phi]=\prod_{i<j} (\Phi_i-\Phi_j)$ being the Vandermonde
determinant.
This formula implies~\cite{KM92} the
following representation of the partition function~\rf{partition}
in terms of the path integral over $\Phi(x)$:
\be
Z_{KM}=\int \prod_x d\Phi(x) \e^{-\sum_x N
\tr{V[\Phi(x)]}} \prod_{x,\mu} I[\Phi(x),\Phi(x+\mu)]\;.
\label{matrixmodel}
\ee
It is instructive to refer \eq{matrixmodel} as the {\it matrix model\/}
representation of the partition function~\rf{partition}.

One can apply now to \eq{matrixmodel} at $N=\infty$ the reduction
prescription described in the previous section which results in
\eq{averages} (or \rf{masterfield}\/) with the reduced models defined by the
formulas like \rf{QRM} for the quenching prescription and or like \rf{TRM}
for the twisting prescription.  However, since the Itzykson--Zuber--Metha
integral depends only on the eigenvalues of $\Phi(x)$ and $\Phi(x+\mu)$, it
does not depend actually on $S(x)$ and $S(x+\mu)$:
\be I[S(x) \Phi
S^\dagger(x), S(x+\mu) \Phi S^\dagger(x+\mu)] = I[\Phi,\Phi]\;.  \label{noS}
\ee
Therefore, the dependence on the quenched momenta or on the matrix
$\Gamma_\mu$ which is constructed from momenta is cancelled.

The averages in the Kazakov--Migdal model can now be calculated according to
\eq{masterfield} via the master field. For the free energy itself, one gets
\be
\frac{1}{Vol.}\log{Z_{KM}}=2\log{(\Delta[\Phi_s])}-N\tr{V[\Phi_s]}
+D\log{I[\Phi_s,\Phi_s]}
\ee
where $Vol.$ stands for the number of sites on the lattice.

Some comments concerning the proposed reduction procedure of the
Kazakov--Migdal model are now in order:
\begin{itemize}
\item While the reduced model involves no explicit dependence on the momenta
entering either $S(x)$ or $\Gamma_\mu$, this dependence survives when
calculating averages of $x$-dependent quantities according to
\eq{masterfield}. The point is that the invariance which remains after the
reduction is solely global $SU(N)$:  $\Phi_s\ra\Omega\Phi_s\Omega^\dagger$,
which can cancel the matrix $S(x)$ at only one point (this corresponds to
the translation invariance of averages in the unreduced
model~\rf{partition}).  \item \eq{noS} is a consequence from the invariance
of the Haar measure in \eq{IZ} under multiplication by a unitary matrix from
the left and from the right separately, \ie $S(x)$ and $S(x+\mu)$ which
appear after the substitution \rf{substitution} can be absorbed by a gauge
transformation.  While only $SU(N)$ gauge transformations are allowed, this
is enough for our purposes since the $U(1)$ part is again cancelled in the
bilinear expressions of the form $[S(x)]_{ij}[S^\dagger(x)]_{kl}$.
\end{itemize}

It is instructive to see of how the planar graphs of the
original model \rf{partition} are reproduced by the proposed reduction
procedure employing the known results for the reduction of scalar
field reported in the previous section.
To this aid, let us calculate the partition function~\rf{partition}
first integrating over scalar field:
 \be
 Z_{KM}=\int \prod_{x,\mu} dU_\mu(x)\,\e^{-S_{ind}[U\mu(x)]}\;,
 \label{gaugefield}
 \ee
 where the {\it induced action\/} for the gauge field
$U_\mu(x)$ is defined by the integral over $\Phi(x)$ in \eq{partition}:
\be
\e^{-S_{ind}[U_\mu(x)]}=\int \prod_x d\Phi(x) \e^{\sum_x N
\tr{\left(-V[\Phi(x)]+
\sum_{\mu}\Phi(x)U_\mu(x)\Phi(x+\mu)U_\mu^\dagger(x)\right)}}\;.
\label{induced}
\ee
I shall refer Eqs. \rf{gaugefield} and \rf{induced} as the {\it gauge
field\/} representation of the Kazakov--Migdal model.

The required planar graphs of the model~\rf{partition} can then be obtained
in two steps. The first one is to calculate the induced action employing the
reduction prescription to the integral
over scalar field on the r.h.s.\ of \eq{induced}. The second step would be
to average over the gauge field according to \eq{gaugefield}. Since I do
not apply the reduction procedure for the gauge field, it would be
enough to show that the two induced actions coincides perturbatively.

For the r.h.s.\ of \eq{induced}, one can apply the formulas of the
previous section which gives for the quenched momentum prescription:
\be
S_{ind}[U_\mu(x)]=
N\sum_{x,\mu}\tr{\left(\Phi_s S^\dagger(x) U_\mu(x) S(x+\mu) \Phi_s
S^\dagger(x+\mu) U_\mu^\dagger(x)S(x) \right)}\;.
\label{indreduced}
\ee
Expanding the matrices $U_\mu(x)$ around the unity and using standard
rules~\cite{Das87} to obtain diagrams for the scalar field, one recovers the
correct diagrammatic expansion for the induced action $S_{ind}[U_\mu(x)]$.

Notice that when one integrates over $U_\mu(x)$ as discussed above, these
diagrams are absorbed into gauge degrees of freedom. This looks precisely
like the mechanism of recovering planar graphs by the master field which has
been proposed in Ref.~\cite{KM92} on the basis on the conjecture about the
translationally invariant master field. I have related therefore this
scenario to the large-$N$ reduction phenomenon and the master field
of Ref.~\cite{KM92} to the saddle point matrix $\Phi_s$.

Up to now I did not discuss what equation determines $\Phi_s$. This
equation can be derived by averaging the (quantum) equation of motion for
$\Phi$ field in the Kazakov--Migdal model, written in the
form~\rf{matrixmodel}, according to the prescription \rf{masterfield}.
Introducing the spectral density $\rho_s(\l)$ which describes the
distribution of eigenvalues of the matrix $\Phi_s$ and using the remaining
global gauge invariance, one gets \be 2 \pint dx
\frac{\rho_s(x)}{\l-x}=\frac{\d V(\l)}{\d\l}- \frac {D}{N^2}\frac{\d}{\d
\l}\left. \frac{\delta \log{I[\rho,\rho_s]}} {\delta \rho(\l)
}\right|_{\rho=\rho_s} \label{mfe} \ee which coincides with the saddle
point equation for the coordinate-independent master field of
Ref.~\cite{KM92}.

I did not presented a direct proof of the reduction prescription
\rf{masterfield} with $\Phi_s$ given by \eq{mfe}.
The above perturbative arguments work only in
the weak coupling region where the perturbative expansion is relevant and
are not applicable in the strong coupling region which is separated
by a phase transition. I hope that loop equations which are derived in
 the next section could be useful for this purpose.

\newsection{Loop equations}

As is pointed out in Ref.~\cite{KhM92}, the simplest gauge invariant objects
in the Kazakov--Migdal model are the {\it adjoint\/} Wilson loops which are
defined by \eq{adjloop} where the
average is understood with the same measure as in \eq{partition}.  The
nonabelian phase factor $U(C_{xx})$ which is associated with a parallel
transport from a point $x$ along a closed loop $C_{xx}$ is defined by the
path ordered product
\be U(C_{xx})=P \prod_{l\in C_{xx}} U_\mu(z)
\label{U}
\ee
where $l$ stands for the link $(z,\mu)$.

One can derive the set of loop equations satisfied by these quantities quite
similarly to the case of standard lattice gauge theory (for a review,
see~\cite{Mig83}) performing an infinitesimal shift
\be
U_\mu(x)\ra \left(1+i\epsilon_\mu(x) U_\mu(x)\right)
\label{shift}
\ee
of the link variable $U_\mu(x)$ at the link $(x,\mu)$ with $\epsilon$ being
an infinitesimal hermitian matrix which leaves the Haar measure invariant.
Since the plaquette term is absent in
the action of the Kazakov--Migdal model, it does not appear on the l.h.s.\
of the loop equation. However, a new term associated with the interaction
between gauge and scalar fields arises on the l.h.s.. The r.h.s. of the loop
equation  satisfied by $W_A(C)$ looks at large $N$ very similar to the
standard one. The equation reads schematically
\bea
\Big\langle \frac 1N
\tr{\left( \Phi(x)U(C_{xx})U_\mu(x)\Phi(x+\mu)U_\mu^\dagger(x)
U^\dagger(C_{xx}) \right)} \Big\rangle &-&\nonumber \\
\Big\langle \frac 1N
\tr{\left( \Phi(x)U^\dagger(C_{xx})U_\mu(x)\Phi(x+\mu)U_\mu^\dagger(x)
U(C_{xx}) \right)} \Big\rangle &=&\nonumber \\
 \sum_{l\in C_{xx}}
\tau_\mu(l) \delta_{xz} W_A(C_{xz}) W_A(C_{zx}) & & \label{le} \eea where
$\tau_\mu(l)$ stands for a unit vector in the direction of the link $l$.

A conceptual difference between \eq{le} and the standard loop equation of
manycolor QCD is that the former one is not closed. While the r.h.s.\ of
\eq{le} involves the same quantity $W_A$, a new gauge invariant object of a
generic type~\rf{G}
emerges on the l.h.s.. One should derive, therefore, an equation satisfied
by this quantity.

The corresponding equation results from the invariance of the measure over
$\Phi$ under an infinitesimal shift
\be
\Phi(x)\ra\Phi(x)+\xi(x)
\label{xi}
\ee
of $\Phi(x)$ at the given site $x$ with $\xi(x)$ being an infinitesimal
hermitian matrix. While to close the set of equations one has to consider
averages of the type~\rf{G} with arbitrary powers of $\Phi(x)$ and
$\Phi(y)$, drastic simplifications occur for the quadratic potential when
the resulting equation reads
\be
2m_0^2 G(C_{xy}) -\sum_{\mu=-D}^D G(C_{(x+\mu)x}C_{xy})
=\delta_{xy} W_A(C_{xy})\;.
\label{sd}
\ee
Here the path $C_{(x+\mu)x}C_{xx}$ is obtained by attaching the link
$(x,\mu)$ to the path $C_{xx}$ at the end point $x$ as is depicted in
Fig.~1.  Notice that $W_A$ for a closed loop enters the r.h.s.\ of \eq{sd}
due to the presence of the delta-function.  Therefore, the set of
Eqs.~\rf{le} and \rf{sd} is closed.

\newsection{Properties of reduced and unreduced averages}

\subsection{Reduced loop averages}

A prescription  to calculate
$W_A(C)$ and $G(C_{xy})$ in the reduced theory can be obtained from
\eq{masterfield}.  The crucial role in this construction is played by the
following integral over the unitary group:  \be
I^{ab}[\Phi,\Psi]=\frac{1}{I[\Phi,\Psi]} \int dU \e^{N\tr{\Phi U \Psi
U^\dagger}}\fr 1N \tr{t^aUt^bU^\dagger}\;, \label{Iab} \ee where $t^a$ for
$a=1,\ldots,N^2-1$ are (hermitian) generators of $SU(N)$ which are normalized
by \be \fr 1N \tr{t^at^b} = \delta^{ab}\;, \hbox{ \ \ \ and \ } [t^a]_{ij}
[t^a]_{kl}=N\delta_{il}\delta_{kj} - \delta_{ij}\delta_{kl}
\label{completeness}
\ee
and $I$ stands for the integral \rf{IZ}. Notice that the $U(1)$-part is
cancelled in this formula so that the integrals over $U(N)$ and $SU(N)$
coincide.

It is convenient to consider $I^{ab}$ as a $(N^2-1)\times(N^2-1)$ matrix and
\be
\Phi^a = \fr 1N \tr{t^a \Phi}
\ee
as a $(N^2-1)$-dimensional vector. Then, the counterparts of $W_A$ and $G$ in
the reduced model read
\be
W_A^{Reduced}=  I^L[\Phi_s,\Phi_s]^{aa}
\label{Wreduced}
\ee
and
\be
G^{Reduced}= \Phi_s^a  I^L[\Phi_s,\Phi_s]^{ab} \Phi_s^b
\label{Greduced}
\ee
where $L$ is the length (in lattice units) of the appropriate contour.

It is easy to see that $W_A^{Reduced}$ and $G^{Reduced}$ given by Eqs.
\rf{Wreduced} and \rf{Greduced} satisfy for an arbitrary potential the same
loop equation \rf{le} as $W_A(C)$ and $G(C_{xy})$ defined by
Eqs.~\rf{adjloop} and \rf{G} in the original unreduced model.
It is trivial to see, employing the representation \rf{Iab} in terms of
the integral over $U$, that the equation coming from the shift \rf{shift}
of $U$ coincides with \eq{le}. The point is that we shift $U$ only in one
integral of the chain (what is an analog of the shift at one link) since one
did not reduce $U$'s (\ie identify them).

The situation is different,
however, for \eq{sd} which results from the shift \rf{xi} of $\Phi$. Now
it is not possible to make any conclusion since the corresponding equation
satisfied by $G^{Reduced}$ can not be derived by a straightforward variation
w.r.t.\ $\Phi_s$. Fortunately \eq{sd} can be exactly solved in the
strong coupling region (see Section~5) and the result can be compared with the
large mass
expansion of $G^{Reduced}$.

\subsection{Large mass expansion}

The above conjecture about the equivalence of the reduced and  original
unreduced models can be tested by comparing the loop averages at large values
of the mass $m_0^2$.
To calculate the large mass expansion, one needs the following expansion of
$I^{ab}$ defined by the integral~\rf{Iab}%
\footnote{The formulas for calculating the corresponding integrals over
the Haar measure as $N\ra\infty$ can be found in Ref.~\cite{Int}.}:
\be
I^{ab}[\Phi,\Psi]=\Phi^a \Psi^b + \ldots \;, \label{expansion} \ee where
$\ldots$ stands for the terms of higher powers in $\Phi$ and $\Psi$ which
correspond to higher order of the large mass expansion.

Applying this formula to the unreduced quantity~\rf{G}, one gets
\be
G(C_{xy})= \prod_{z\in C_{xy}} \left(
\frac{\int d\Phi(z)\e^{-m_0^2N\tr{\Phi^2(z)}}
  \fr 1N \tr{\Phi^2(z)}}{\int d\Phi(z)\e^{-m_0^2N\tr{\Phi^2(z)}}}
\right)+\ldots = (G_0)^{L+1}+ \ldots \label{Glo} \ee
while the reduced prescription~\rf{Greduced} gives
\be
G^{Reduced}=\left(\fr 1N \tr{\Phi_s^2} \right)^{L+1}+\ldots
\label{Greducedlo}
\ee
which coincides with \rf{Glo} since $\fr 1N
\tr{\Phi_s^2}=G_0$. Therefore, one has explicitly demonstrated, in
particular, the reduction to the leading order of the large mass  expansion.

A similar analyses of the adjoint Wilson loops leads to a different result.
Let us now estimate the average on the r.h.s.\ of \eq{adjloop} for a
closed loop with $L\geq 4$ which does not contain parts that are passed back
and forth.  Exploiting the completeness condition~\rf{completeness}
and \eq{expansion}, the
result can be represented in the same form as~\rf{Glo} but with an extra
factor $1/N^2$:  \be W_A(C)= \frac{1}{N^2} (G_0)^L + \ldots \;.  \label{Wlo}
\ee This expression is proportional to $1/N^2$ and vanishes in the large-$N$
limit which is in agreement with general arguments of Ref.~\cite{KhM92}.
The corresponding average in the reduced model can be calculated similarly
to \eq{Greducedlo}. The result is given by the r.h.s.\ of \eq{Wlo} and now
holds independently of whether the loop is closed or open.

\subsection{Intermediate coupling region}

When $m_0^2$ is decreased, the system undergoes phase transitions. According
to the scenario of Ref.~\cite{KhM92} which is discussed in Section~1, a
first order large-$N$ phase transition should occur before the one
associated with the continuum limit in order for the Kazakov--Migdal model
to induce continuum QCD. While the strong and weak coupling phases always
exist (the weak coupling phase is associated as usual with the perturbative
expansion) the existence of such an {\it intermediate} phase has been only
conjectured. Let us discuss consequences from the phenomenon of large-$N$
reduction in the intermediate coupling region.

In the weak coupling region where the perturbative expansion is applicable
the arguments of Section~2.2 about the large-$N$ reduction work.

The intermediate coupling region looks pretty similar to the weak coupling
phase of Wilson lattice gauge theory.
My arguments are based on the gauge field
representation~\rf{gaugefield} of the Kazakov--Migdal model. Let us
represent the induced action in the form
\be
S_{ind}[U]=-\frac 12 \sum_{C} \beta_A(C) \left|\tr{U(C)}\right|^2
\label{ind}
\ee
where $\beta_A(C)$ are some loop-dependent couplings which are determined
by the potential $V[\Phi]$. For the quadratic potential, one gets~\cite{KM92}
\be
\beta_A(C)=\frac{1}{l(C)m_0^{2l(C)}}
\label{qind}
\ee
with $l(C)$ being the length of the loop $C$.

At $N=\infty$ the action
\rf{ind} is equivalent to the following action
\be
S_{F}[U]=-N \sum_C \bar{\beta}(C) \Re \hbox{\,tr\,}{U(C)}
\label{fund}
\ee
with the couplings $\bb(C)$ being determined by the self-consistency
conditions
\be
\bb(C)=\beta_A(C) W_F(C;\{\bb\})
\label{self-consistency}
\ee
where $W_F(C;\{\bb\})$ is the {\it fundamental} Wilson loop average
\be
W_F(C;\{\bb\}) =\left\langle \frac 1N \tr{U(C)} \right\rangle
\ee
and the averaging is taken with the action \rf{fund}. This procedure extends
the one advocated by Khokhlachev and the author~\cite{KhM81} for the
single-plaquette adjoint action
\be
S_A = -\frac{\beta_A}{2} \sum_\Box \left|\tr{U(\Box)}\right|^2
\label{adjaction}
\ee
with the self-consistency condition given by \eq{self-consistency} with
$C=\Box$.

In the intermediate coupling region
Eqs.~\rf{self-consistency} possess a nontrivial solution and the model can
be described in terms of the fundamental induced action \rf{fund}.
On the other hand since the area law holds for the Wilson loop averages, I
expect that the contribution of long loops to the induced action~\rf{fund}
is suppressed according to \eq{self-consistency}. Therefore, properties of
the intermediate coupling region should be similar to the standard lattice
gauge theory.

\newsection{An exact strong coupling solution}

\subsection{Quadratic potential}

A drastic simplification of the loop equations \rf{le} and \rf{sd} is due to
the fact~\cite{KhM92} that in the strong coupling region (before the
large-$N$ phase transition) the adjoint Wilson loops vanish in the large-$N$
limit except closed ones with vanishing area $A_{min}(C)$  of the minimal
surface (\ie {\it contractable} to a point owing to unitarity of $U$'s): \be
W_A(C)=\delta_{0A_{min}(C)}+{\cal O}\left({1\over N^2} \right) \;.
\label{1overN} \ee
It can be shown that \rf{1overN} is consistent with loop equations.

Given this behavior of adjoint Wilson loops
entering the r.h.s.\ of \eq{sd}, \eq{sd}
admits a solution $G(C_{xy})=G_L$ depending on the single parameter $L$
which is defined as the {\it algebraic} length of the loop
(see~\cite{Mig83}), \ie the length after all possible contractions of the
paths passing back and forth are made. For such an ansatz, \eq{sd} can
be written as
\bea
2m_0^2 G_{L} - G_{L-1} - (2D-1)G_{L+1}& =& 0 \hbox{ \ \ \
for \ \ } L\geq1\;, \nonumber \\ 2m_0^2 G_0 - 2D G_1&=&1\;.  \label{ansatz}
\eea

Before solving this equation let me show of how it is satisfied in $D=1$
where the exact result for $G_L$ is given by the Laplace transform
\be
G_L^{D=1}=\frac 12 \int_0^\infty d\alpha \e^{-m_0^2
\alpha}\,\hbox{I}_L(\alpha)
=\frac{\left(m_0^2-\sqrt{m_0^4-1}\right)^L}{2\sqrt{m_0^4-1}}
\label{D=1}
\ee
with \ I$_L(\alpha)$ being modified Bessel functions. \eq{ansatz} can now be
viewed, say, as the recurrent relation
\be
2\hbox{I}_L^\prime(\alpha)-\hbox{I}_{L-1}(\alpha)-\hbox{I}_{L+1}(\alpha)=0
\ee
which is integrated according to \eq{D=1} while $1$ on the r.h.s.\ of the
$L=0$ equation results from the fact that \ I$_0(0)=1$.

\eq{ansatz} can be solved by introducing the generating function
\be G(\l)=\sum_{L=0}^\infty
G_L \l^L \ee
with the result being expressed via the initial data, $G_0$, by
\be
G(\l)=\frac{\fr 1D \left( m_0^2G_0+D-\half \right)\l-(2D-1)G_0
}{2m_0^2\l-\l^2+1-2D}.  \label{solution} \ee

While $G_0$ is defined as an average of $\fr 1N \tr{\Phi^2(x)}$ in the
Kazakov--Migdal model, one can determine it from \eq{solution} by imposing
the analytic structure of $G(\l)$ as a function of the spectral parameter
$\l$. For an arbitrary $G_0$ the denominator in \eq{solution} has two roots
and $G(\l)$ has therefore two poles. The exact solution at $D=1$ given by
\eq{D=1} as well as the large mass expansion  described in Section 4.2 lead
to a $G(\l)$ possessing only one pole. This can be achieved in \eq{solution}
by choosing $G_0$ in a proper way. This requirement unambiguously determines
$G(\l)$ to be
\be
G(\l)=\frac{(2D-1)^2}{2\left( m_0^2(D-1)+D\sqrt{m_0^4+1-2D}\right)
\left[\left(\sqrt{m_0^4+1-2D}-m_0^2\right) \l +2D-1 \right]}
\label{final}
\ee
while $G_0$ is fixed to be
\be
G_0=\frac{D-\half}{m_0^2(D-1)+D\sqrt{m_0^4+1-2D}}\;.
\label{gross}
\ee
I have checked that the solution \rf{final}
correctly reproduces the $D=1$ solution \rf{D=1} as well as the leading
order of the large mass expansion \rf{Glo} at any $D$.

Some comments about the exact strong coupling solution \rf{final} are now in
order:
\begin{itemize}
\item \eq{gross} precisely coincides with the result by Gross~\cite{Gro92}
obtained by another method.
\item
While the solution \rf{final} looks very simple, it
corresponds in the language of the Kazakov--Migdal model to tedious
calculations of integrals over unitary matrices with a subsequent averaging
over $\Phi$. In the reduced language \rf{Greduced} it remains still to
calculate the integral \rf{Iab} over unitary group and then substitute the
value of the master field $\Phi_s$. It would be very interesting to
reproduce the result by this method.
\item
The solution \rf{final} is the first example of calculations of {\it
extended\/} objects (since $G_L$ is the average \rf{G} for the loop of the
length $L$ in the lattice units) in the strong coupling region of the
Kazakov--Migdal model while the results of Refs.\cite{KM92,Mig92a,Gro92}
refer to {\it local\/} objects like the spectral density.
\item
The very possibility to find such a simple exact solution to loop equations
in the strong coupling region is related to a very simple form \rf{1overN}
of the Wilson loops. In particular, while the path $C_{xy}$ looks like a
string in the strong coupling expansion of the standard lattice gauge
theory, fluctuations of its shape are now suppressed. This is why the result
depends only on $L$.
\item The solution \rf{final} at $D=1$ could be interesting from the
viewpoint of matrix models of 2D gravity.  \end{itemize}

\subsection{More general potentials}

While the solution~\rf{final} has been obtained for the quadratic potential,
an analogous solution can be found for the more general potential
 \be
 N\tr{V[\Phi]}=m_0^2 N\tr{\Phi^2} +g(\tr{\Phi^2})^2
 \label{quartic}
 \ee
where $g\sim1$ as $N\ra\infty$ to provide self-interaction of scalar field.
 This kind of potential was used~\cite{AGB92} in the matrix models in the
 context of $D>1$ strings.

The Kazakov--Migdal model with the quartic potential~\rf{quartic} can be
solved in the large-$N$ limit for the following reasons. The model with
the potential~\rf{quartic} is equivalent as $N\ra \infty$ to the one with
 quadratic potential whose mass parameter ${\bm}^2$ is defined by the
 self-consistency relation \be {\bm}^2 = m_0^2 +2 g\left.
 G_0\right|_{{\bm}^2} \label{mbar} \ee where $\left. G_0\right|_{{\bm}^2}$
  is given by \eq{gross} with $m_0^2$ replaced by ${\bm}^2$. \eq{mbar} can
  be obtained naively replacing one $\tr{\Phi^2}$ by the average value due
  to factorization. A rigorous proof of \eq{mbar} can be done using loop
  equations similarly to the proof~\cite{KhM81} of the reduction of the
  adjoint action to the Wilson action which is discussed in Section~4.3.

  \eq{mbar} can be used to study whether the large-$N$ phase transition
  occurs for the potential~\rf{quartic}. Such a phase transition were occur
  if ${\bm}^2$ would depend on $m_0^2$ nonmonotonically. A similar idea for
  the large-$N$ phase transition  to occur in lattice gauge theories was
  advocated in Ref.~\cite{MP81}. For the scalar model of the
  type~\rf{quartic}, it was employed in Ref.~\cite{AGB92} to obtain
  $\gamma_{string}>0$.

  Given Eqs. \rf{mbar} and \rf{gross} it is easy to calculate the derivative
  \be
  \frac{\d m_0^2}{\d {\bm}^2}
  =1+\frac{g(2D-1)\left(D-1+\frac{{\bm}^2
  D}{\sqrt{{\bm}^4+1-2D}}\right)}{\left(
  \bm^2(D-1)+D\sqrt{\bar{m}^4+1-2D}\right)^2}
  \label{partial}
  \ee
  and see that it is positive for ${\bm}^2 >D$ where the gaussian model is
  stable. Therefore I conclude that, similarly to the case of the quadratic
  potential, {\it there is no first order large-$N$ phase transition} for
  the Kazakov--Migdal model with the potential~\rf{quartic}.

  An analogous study can be performed for the potential~\rf{general}
  with an arbitrary function $f$ when the second term on the r.h.s.\ of
  \eq{partial} is multiplied by $\fr 12 f^{\prime\prime}(G_0)$. A conclusion
  is that one has to look for a more complicated potential.

\section*{Acknowledgments}

I am grateful to S.Khokhlachev for useful discussions and suggestions.

\eop

\eop

\section*{Figures}
\vspace{4cm}
\unitlength=1.00mm
\linethickness{0.4pt}
\begin{picture}(141.00,90.00)
\put(30.00,50.00){\line(3,0){20.00}}
\put(50.00,90.00){\line(3,0){20.00}}
\put(100.00,50.00){\line(3,0){20.00}}
\put(120.00,90.00){\line(3,0){20.00}}
\put(30.00,55.00){\makebox(0,0)[cc]{$x$}}
\put(70.00,83.00){\makebox(0,0)[cc]{$y$}}
\put(100.00,55.00){\makebox(0,0)[cc]{$x$}}
\put(140.00,83.00){\makebox(0,0)[cc]{$y$}}
\put(90.00,44.00){\line(5,3){10.00}}
\put(90.00,37.00){\makebox(0,0)[cc]{$x+\mu$}}
\put(50.00,20.00){\makebox(0,0)[cc]{\large{a)}}}
\put(120.00,20.00){\makebox(0,0)[cc]{{\large b)}}}
\put(50.00,50.00){\vector(0,1){20.00}}
\put(50.00,90.00){\line(0,-3){20.00}}
\put(120.00,50.00){\vector(0,1){20.00}}
\put(120.00,70.00){\line(0,3){20.00}}
\put(30.00,49.00){\circle*{2.00}}
\put(70.00,89.00){\circle*{2.00}}
\put(52.00,88.00){\vector(0,-1){20.00}}
\put(52.00,48.00){\line(0,3){20.00}}
\put(30.00,48.00){\line(3,0){22.00}}
\put(52.00,88.00){\line(3,0){18.00}}
\put(90.00,43.00){\circle*{2.00}}
\put(140.00,89.00){\circle*{2.00}}
\put(122.00,88.00){\line(3,0){18.00}}
\put(122.00,88.00){\vector(0,-1){20.00}}
\put(122.00,48.00){\line(0,3){20.00}}
\put(100.00,48.00){\line(3,0){22.00}}
\put(90.00,42.00){\line(5,3){10.00}}
\end{picture}

\vspace{2cm}

\begin{description}
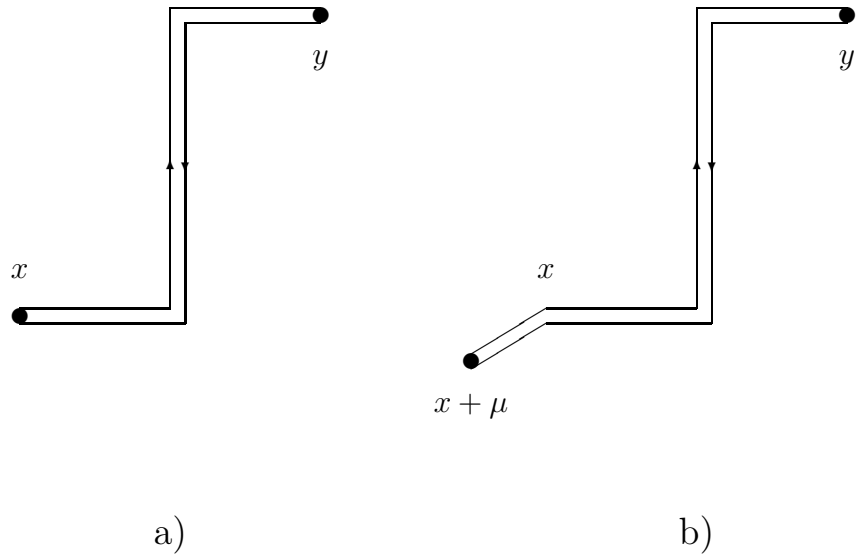
   \item[Fig. 1] \ \ \ The graphic representation for $\Phi(C_{xy})$ (a)
   and $\Phi(C_{(x+\mu)x}C_{xy})$ (b) entering \eq{sd}. The bold points
   represent $\Phi(x)$ and $\Phi(x+\mu)$. The (oriented) solid lines
   represent the path-ordered products $U(C_{xy})$ and $U(C_{(x+\mu)x}C_{xy})$.
   The color indices are contracted according to the arrows.
\end{description}

\end{document}